\documentclass[journal]{IEEEtran}
\pdfoutput=1

\usepackage{ifpdf}
\usepackage{subfigure}
\usepackage{graphicx}
\DeclareGraphicsExtensions{.pdf,.jpg,.png}
\usepackage[cmex10]{amsmath}
\usepackage{amssymb}
\usepackage{array}
\usepackage{mdwmath}
\usepackage{mdwtab}
\usepackage{pslatex}
\usepackage{xcolor}
\usepackage{textcomp}
\usepackage{placeins}
\usepackage{eqparbox}
\usepackage{url}
\usepackage{stfloats}
\usepackage{multicol}
\usepackage{subfigure}
\usepackage{float}
\usepackage{algorithmicx}
\usepackage{algpseudocode}
\usepackage{amsthm}
\usepackage{comment}
\usepackage{balance}

\begin{document}

\title{IEEE 802.15.3d: First Standardization Efforts for Sub-Terahertz Band Communications towards 6G}

\author{Vitaly Petrov, Thomas Kürner, and Iwao Hosako
\thanks{Vitaly~Petrov is with Tampere University, Finland. This work was partially completed when staying with Technische Universität Braunschweig, Germany.}
\thanks{Thomas Kürner is with Technische Universität Braunschweig, Germany and also has been serving as a chair of the IEEE 802.15.3d Task Group.}
\thanks{Iwao Hosako is with the National Institute of Information and Communications Technology (NICT), Japan and also has been serving as a vice-chair of the IEEE 802.13.3d Task Group.}
}

\maketitle

\begin{abstract}
With the ratification of the IEEE 802.15.3d amendment to the 802.15.3, a first step has been made to standardize consumer wireless communications in the sub-THz frequency band. The IEEE 802.15.3d offers switched point-to-point connectivity with the data rates of 100\,Gbit/s and higher at distances ranging from tens of centimeters up to a few hundred meters. In this article, we provide a detailed introduction to the IEEE 802.15.3d and the key design principles beyond the developed standard. We particularly describe the target applications and usage scenarios, as well as the specifics of the IEEE 802.15.3d physical and medium access layers. Later, we present the results of the initial performance evaluation of IEEE 802.15.3d wireless communications. The obtained first-order performance predictions show non-incremental benefits compared to the characteristics of the fifth-generation wireless systems, thus paving the way towards the six-generation (6G) THz networks. We conclude the article by outlining the further standardization and regulatory activities on wireless networking in the THz frequency band.
\end{abstract}

\section{Introduction}
\label{sect:intro}

Directional wireless communications over millimeter-wave (mmWave, $30$\,GHz--$300$\,GHz) bands is one of the most significant novelties introduced in the fifth-generation (5G) wireless networks. Adopted by 3rd Generation Partnership Project (3GPP) for New Radio (NR) technology, mmWave communications at the frequencies between $24.25$\,GHz and $52.6$\,GHz allow for the data exchange at the rates of \emph{several gigabits per second (Gbps)}. The initial phase of the standardization process for mmWave NR was finished in 2018, and the first deployments already started to appear in Europe, Japan, the United Kingdom, and the United States.

Along this timeline, the utilization of mmWave communications at $60$\,GHz has also been ratified by the Institute of Electrical and Electronics Engineers (IEEE) originally for wireless transmission of high-definition video in the IEEE Std. 802.15.3c--2009 and then for wireless local area networks (WLANs) in the IEEE Std. 802.11ad--2012. Depending on the country-specific regulations, up to $6$ channels of $2.16$\,GHz each can be supported in IEEE 802.11ad terminals. The IEEE is currently working on further mmWave-specific features in IEEE P802.11ay, aiming to finalize it by the end of 2020.

The development of 5G-grade mobile mmWave communications is now almost complete, and vendors are testing their early implementations. The research focus is thus shifting towards determining the vision, requirements, and technology enablers for future wireless networks~\cite{beyond_5g_vision}. Due to the complexity in adopting mmWave in 5G, many stakeholders agree that the next-generation systems (often termed as 5G+) will primarily improve the efficiency of data exchange over existing $28$--$73$\,GHz bands. Simultaneously, harnessing the frequencies above $100$\,GHz is planned for 6G and beyond solutions as a key enabler for new applications demanding \emph{ultra-high user data rates exceeding tens of gigabits per second}~\cite{oulu_6g}.

The 6G vision is currently at the very early stages of its development. The support for futuristic scenarios is particularly considered, such as ubiquitous penetration of augmented and virtual reality (AR/VR) systems, holographic telepresence, and collective driving by autonomous robots~\cite{rangan_6G_use_cases_arxiv_commag}. From the technology perspective, the prospective 6G networks are also envisioned to introduce a number of novelties. These include full-duplex wireless communications, exploitation of machine learning for performance optimization, multi-radio multi-connectivity solutions, and many more~\cite{rap_6g}.

Concerning the extreme data rates to support, 6G systems are expected to complement microwave and mmWave connectivity options with wireless communications over the frequencies far above $100$\,GHz, particularly the Terahertz (THz) band communications ($300$\,GHz--$3$\,THz) and Visible Light Communications (VLC, $400$\,THz--$800$\,THz), both providing \emph{continuous bands of tens, hundreds, or even thousands of gigahertz}. The major players from academia, industry, standardization, and regulatory bodies have been actively exploring the feasibility of such communications over the last decades. For VLC, this effort has recently resulted in the creation of the IEEE Std. 802.15.7--2018 standard summarizing the physical (PHY) layer and the medium access control (MAC) sublayer for prospective optical wireless communications.

Standardization of THz wireless communications by the IEEE started in early 2008 when a ``Terahertz Interest Group'' (IGthz) has been established under the 802.15 umbrella. The key design choices and preliminary performance predictions made by the group members laid down the foundation of the IEEE Task Group on ``100G Wireless'' (TG100G, IEEE 802.15.3d) in 2014. The first IEEE standard for wireless communications over the low THz band (or \emph{sub-THz band}) by TG100G, IEEE Std. 802.15.3d--2017, has been officially approved in Fall 2017. While the key contributors are currently working on their prototype solutions by directly following the specifications~\cite{ieee_802_15_3d}, the successful adoption of the standard requires a better understanding of its main features and the existing limitations by a broader research community. The latter brings us to the main goal of this article.

In this paper, we present an overview of the IEEE Std. 802.15.3d--2017 -- the first IEEE-family standard for wireless communications over up to 69\,GHz-wide channels in the sub-THz band (specifically, $253$\,GHz--$322$\,GHz). We first review the target applications and usage scenarios in Section~\ref{sec:applications}. We then present the major details of the MAC and PHY layers in Section~\ref{sec:mac} and Section~\ref{sec:phy}, respectively. The initial performance assessment is reported in Section~\ref{sec:performance}. We finally outline the future work for the development and standardization of THz band communications in Section~\ref{sec:future}. 

\section{IEEE~802.15.3d Motivation and Usage Scenarios}
\label{sec:applications}

\begin{figure}[!t]
 \centering
 \includegraphics[width=0.9\columnwidth]{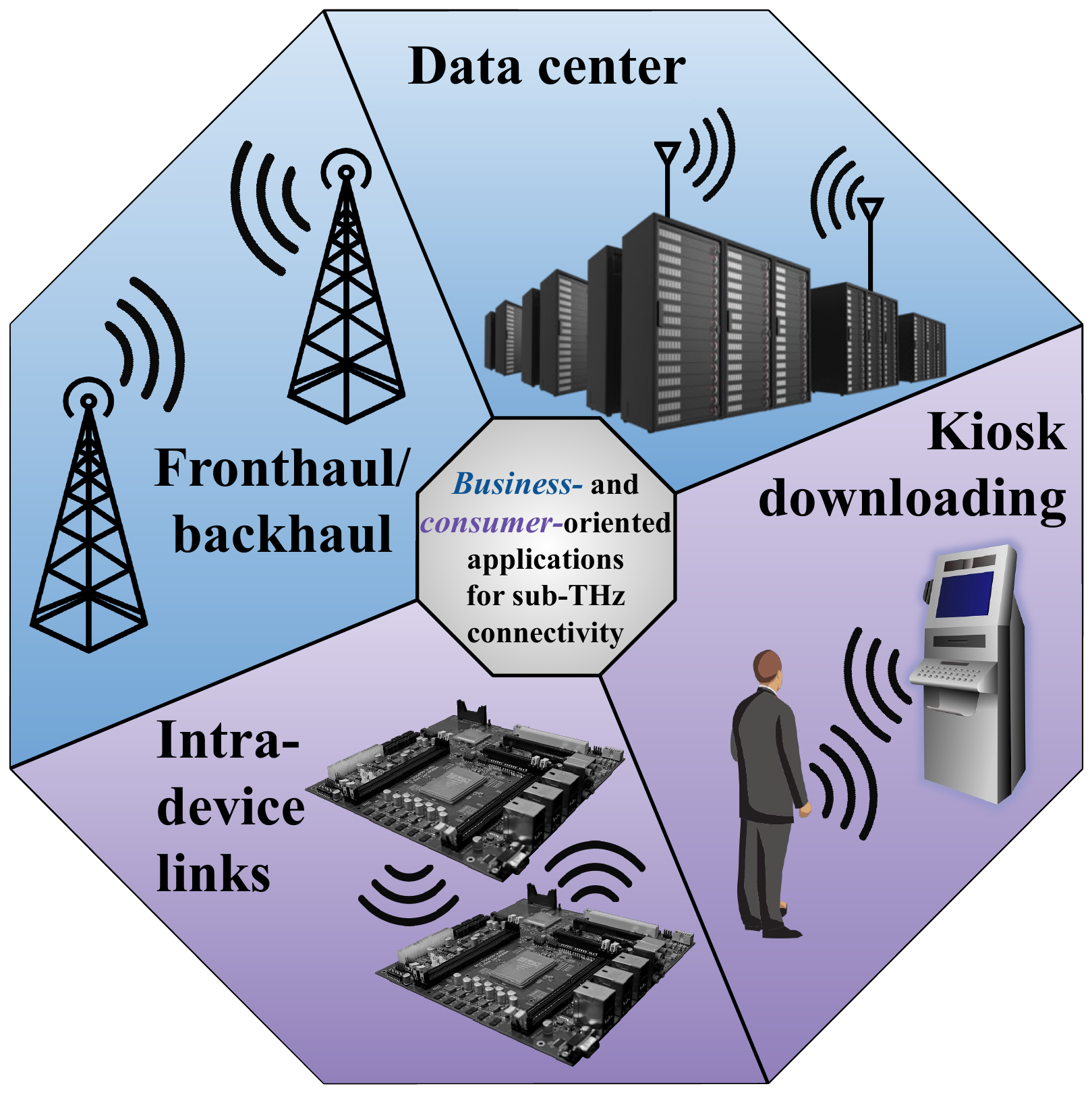}
 \caption{Target applications for the IEEE 802.15.3d.}
 \label{fig:applications}
\end{figure}

The introduced IEEE Std. 802.15.3--2017 is an amendment to the IEEE Std. 802.15.3--2016. It defines a wireless PHY layer operating at $100$\,Gbit/s with a possible fallback to lower rates. The main objective is to illustrate the feasibility of sub-THz communications, as well as to provide a standard for low-complexity and high-rate connectivity. The rates are high enough for the emerging applications (see Fig.~\ref{fig:applications}), outlined further in this section in decreasing communication range.

The IEEE Std. 802.15.3d--2017, as one of the first standardization efforts for wireless communications at $300$\,GHz, primarily targets the applications that are feasible with the current level of electronics. Notably, the considered deployments are limited to point-to-point links between the devices that remain static or quasi-static during the data exchange~\cite{thz_ard}. The latter gives the key difference with the IEEE 802.11-family standards (particularly, IEEE Std. 802.11ad--2014 and IEEE P802.11ay) that primarily target mobile point-to-multipoint communications at $60$\,GHz with lower data rates.

The stationary and inherently point-to-point nature of the links allows reducing the complexity of the IEEE Std. 802.15.3d--2017 MAC since the requirements for multiple access and interference mitigation are relaxed. In addition, simpler procedures for initial access and device discovery can be used. Apart from the lower complexity in the implementation, the signaling overhead is reduced, allowing higher net data rates. Support of nodes mobility, as well as multiple channel access, is investigated for the future work~\cite{petrov_wcm_radar}.

\subsection{Wireless Fronthaul and Backhaul Links}
Future wireless networks are envisioned to utilize ultra-dense deployments of mmWave small cells to satisfy the projected capacity demands in crowded areas. One of the challenges here is to provide a reliable and high-rate \emph{backhaul} connection between the mmWave cell and the core network. In addition, when a small cell is equipped with several remote radio heads (RRHs), the stable and fast \emph{fronthaul} connection is needed between the RRHs and the base band unit (BBU).

Here, the massive utilization of optical fiber connections increases the ``transportation costs'' and complicates the deployment of mmWave small cells. Therefore, a wireless alternative has been proposed to complement the optical links in challenging setups. The approach is referred to as \emph{Integrated Access and Backhaul (IAB)} and suggests reusing the mmWave access technology for the backhaul~\cite{iab}. This solution is flexible but suffers from the limited data rate. The introduced IEEE Std. 802.15.3d--2017 combines the strengths of both options. On one side, the standard supports the data rates of $\approx100$\,Gbit/s -- that is an order of magnitude higher than the state-of-the-art mmWave technologies. On the other side, the use of IEEE Std. 802.15.3d--2017 notably simplifies the cabling.

\subsection{Additional Wireless Links in Data Centers}

Directional links via the IEEE Std. 802.15.3d--2017 can also be applied for wireless connections between the server racks in future data centers. One of the bottlenecks here is enabling efficient data exchange among the computing and storage nodes. Current solutions primarily utilize high-speed Ethernet via a fiber optic. However, the need for many cable channels challenges the reconfigurability and complicates cooling since some air streams become blocked.

Today, an alternative solution is actively studied, suggesting to utilize high-rate point-to-point wireless links when connecting the racks with each other. This approach enables a more flexible design of the data centers as well as reduces the amount of cabling. The IEEE Std. 802.15.3d--2017 aims to provide this ``inter-rack'' connectivity, thus complementing the fiber optics in certain deployments~\cite{thz_ard}.

\subsection{Kiosk Downloads}
Moving forward from business-oriented to consumer usage scenarios, the IEEE Std. 802.15.3d--2017 aims to offer high-rate data exchange for close-proximity communications. The two subcategories here are: (i) device-to-device (D2D) communications and (ii) kiosk download (see Fig.~\ref{fig:applications}).

The standardized sub-THz link offers an order of magnitude greater capacity in comparison with existing and emerging close-proximity wireless solutions in microwave or even mmWave bands. Particularly, the download time of a 2-hours-long movie ($900$\,MB) gets reduced from around $10$\,s with the IEEE 802.11ac to $0.1$\,s with the IEEE Std. 802.15.3d--2017~\cite{thz_ard}.

\subsection{Intra-Device Communications}
Further narrowing the range, not only consumer devices but also the electronic components inside a single device may be connected with the IEEE Std. 802.15.3d--2017. Today, modern computers are already equipped with several high-rate wired links interconnecting, e.g., the central processing unit (CPU) to the random-access memory (RAM), chipset to the network interface, and more. In addition, extremely-complex solutions are utilized in emerging massive multi-core CPUs to connect the computing cores and the shared cache memory.

The rapid growth in the data exchange inside the computers challenges their design. For instance, modern motherboards already have up to $12$ layers, while the emerging ``networks-on-chip'' for the future CPUs occupy over $30$\% of the processor space and power consumption. Here, the IEEE Std. 802.15.3d--2017 provides a possible alternative, enabling the high-rate data links between the critical components, while simultaneously simplifying the layout design~\cite{thz_ard,sergi_wnoc}.

\section{IEEE~802.15.3d MAC Layer}
\label{sec:mac}

\begin{figure}[!t]
  \centering
  \includegraphics[width=0.95\columnwidth]{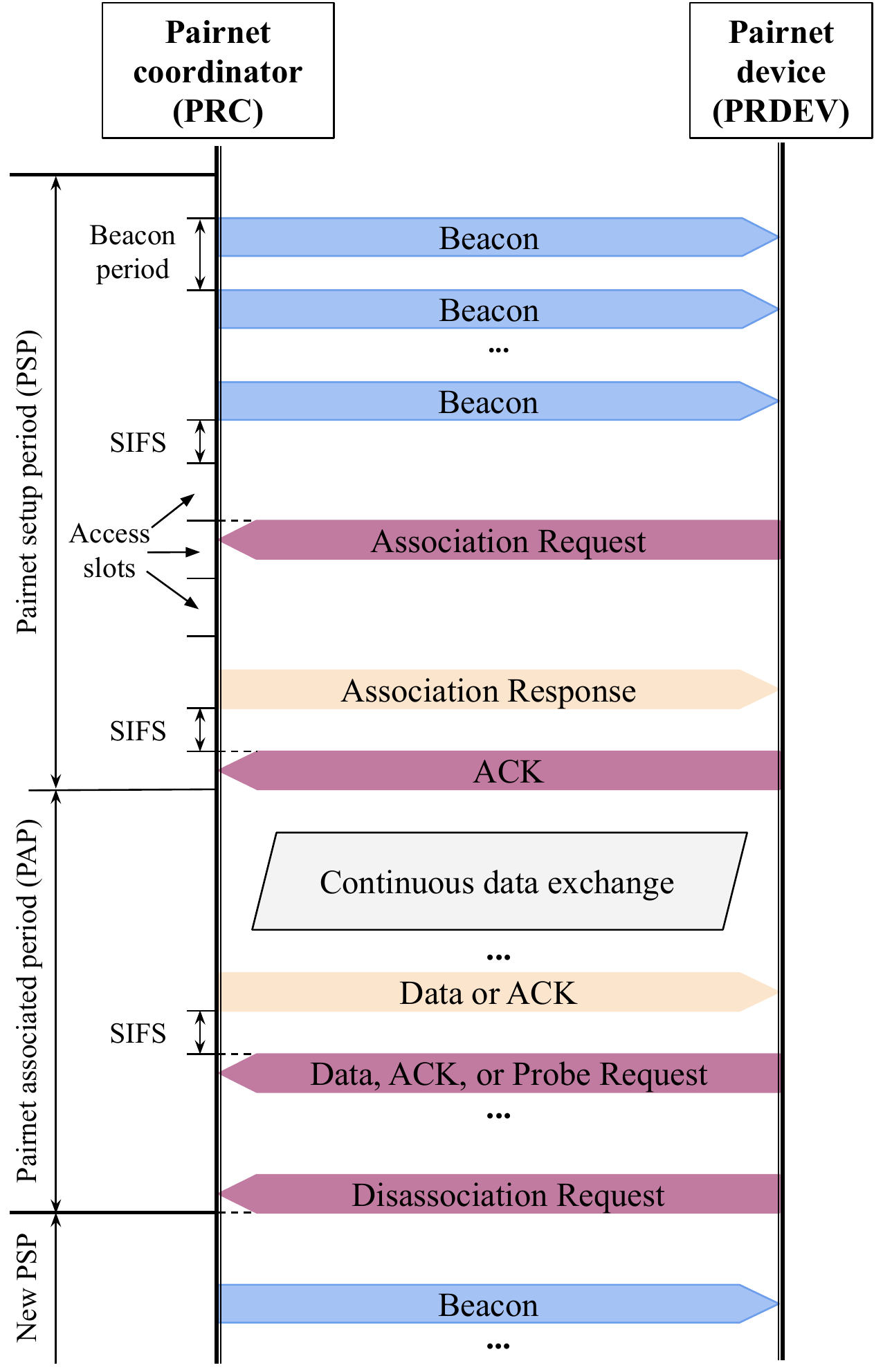}
  \caption{Pairnet MAC layer signaling used in the IEEE 802.15.3d.}
  \label{fig:mac_signaling}
\end{figure}

In contrast to IEEE 802.11 and other standards for WLANs, IEEE Std. 802.15.3d--2017 support only point-to-point communications. The standard follows the MAC layer defined in the IEEE Std. 802.15.3e--2017~\cite{ieee_802_15_3e} and thus utilizes the concept of a \emph{pairnet} connecting no more than two devices. The point-to-point nature of the communications limits the range of possible use cases, but, at the same time, reduces the problem of interference and ``fighting for access''. The high-level description of the major signaling is presented in Fig.~\ref{fig:mac_signaling}.

The communications are divided into two major periods:
\begin{enumerate}
\item Pairnet setup period (PSP),
\item Pairnet associated period (PAP).
\end{enumerate}

During the PSP, one of the devices termed as the \emph{pairnet coordinator (PRC)} creates the pairnet and starts periodically sending \emph{Beacons} with the network details. The beacon frame includes the information on the number and duration of the access slots that can be used by the device willing to join the pairnet. Then, another device ready to joint the pairnet (termed as the \emph{pairnet device (PRDEV)}) responds to the beacon by sending the \emph{Association Request} at the beginning of one of the defined access slots. After a successful reception of the Association Request, the PRC stops transmitting beacons and sends an \emph{Association Response} to the PRDEV, ending the PSP. In addition to the pairnet setup, higher layer protocol setups, e.g., Internet Protocol (IP) or object exchange file transfer (OBEX), may be performed during the PSP. These additions are supported as both Association Request and Association Response frames have the fields to include higher-layer protocol information.

The second period -- PAP -- starts with the successful reception of the Association Response and is dedicated to the actual data exchange. During this period, both the PRC and the PRDEV transmit data frames and optional acknowledgments, starting with the PRDEV. The frames transmitted during PAP are separated by the defined \emph{short interframe space (SIFS)}. When one of the nodes wants to terminate the communications, it sends the \emph{Disassociation Request}. The PRC may also end PAP if there was no message received from the PRDEV during the defined timeout. In case the PRDEV does not want to terminate the communications but has no actual data, it may transmit a \emph{Probe Request} to restart the PRC timeout timer. Whenever the PAP is over (either via a Disassociation Request or by a timeout), the PRC switches back to the PSP: continues transmitting beacons and waiting for the new connections.

\section{IEEE~802.15.3d PHY Layer}
\label{sec:phy}

\subsection{General Overview}
\label{sec:phy_general}

The PHY layer in the IEEE Std. 802.15.3d--2017 has two modes:
\begin{itemize}
\item THz single carrier mode (\emph{THz-SC PHY}), as described in Subsection~\ref{sec:phy_sc},
\item THz on-off keying mode (\emph{THz-OOK PHY}), as described in Subsection~\ref{sec:phy_ook}.
\end{itemize}

The first of these PHY modes, THz-SC PHY, is designed for the high data rates. This mode primarily targets the bandwidth-oriented use cases from Section~\ref{sec:applications}, such as wireless fronthaul/backhaul and additional links in the data center. On its turn, the second mode, THz-OOK PHY, targets the lower-cost sub-THz devices that are not able to utilize complex signals and thus have to rely on the amplitude information. This is the case, for example, with solutions employing resonant tunneling diodes (RTD). Still, the rates of tens of Gbit/s can be achieved with the THz-OOK PHY when the widest channels are used.

\subsection{Supported Channels}
\label{sec:channels}
\begin{figure}[!h]
 \centering
 \includegraphics[width=1.0\columnwidth]{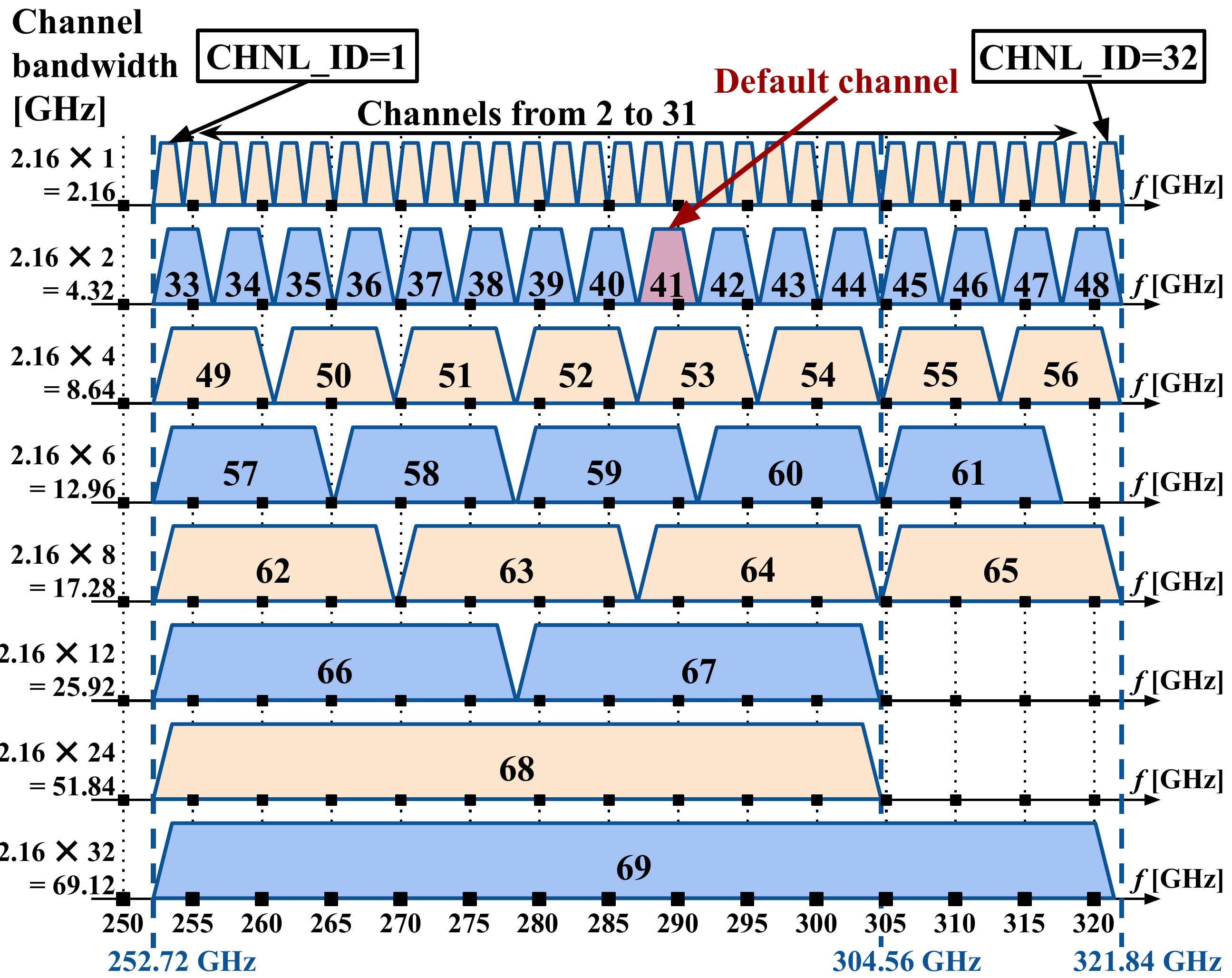}
 \caption{IEEE 802.15.3d channel plan.}
 \label{fig:channels}
\end{figure}

The standard may operate over the sub-THz frequencies between $252.72$\,GHz and $321.84$\,GHz. In total, there are $69$ overlapping channels (see Fig.~\ref{fig:channels}). There are $8$ supported channel bandwidths ranging from $2.16$\,GHz, as in mmWave IEEE Std. 802.11ad--2012, and up to massive $69$\,GHz. Depending on the use case and the hardware capabilities, either the entire frequency range can be allocated for a single $69.12$\,GHz-wide channel (CHNL\_ID=69) or shared between several smaller channels. The bandwidths of the smaller channels are integer multiples of $2.16$ GHz. The channel number $41$ with the bandwidth of $4.32$\,GHz ($2\times2.16$\,GHz) is the default channel.

Following the decision at World Radio Conference 2019 (WRC-2019), all the bands depicted in Fig.~\ref{fig:channels} are available for THz communications globally if specific conditions to protect radio astronomy and earth exploration satellite service are met, as further discussed in Section~\ref{subsec:regulatory}. These conditions do not explicitly specify any transmit power limits and are applicable, in practice, primarily to the narrow areas surrounding the ground radio astronomy stations.

\begin{figure}[!t]
 \centering
 \includegraphics[width=1.0\columnwidth]{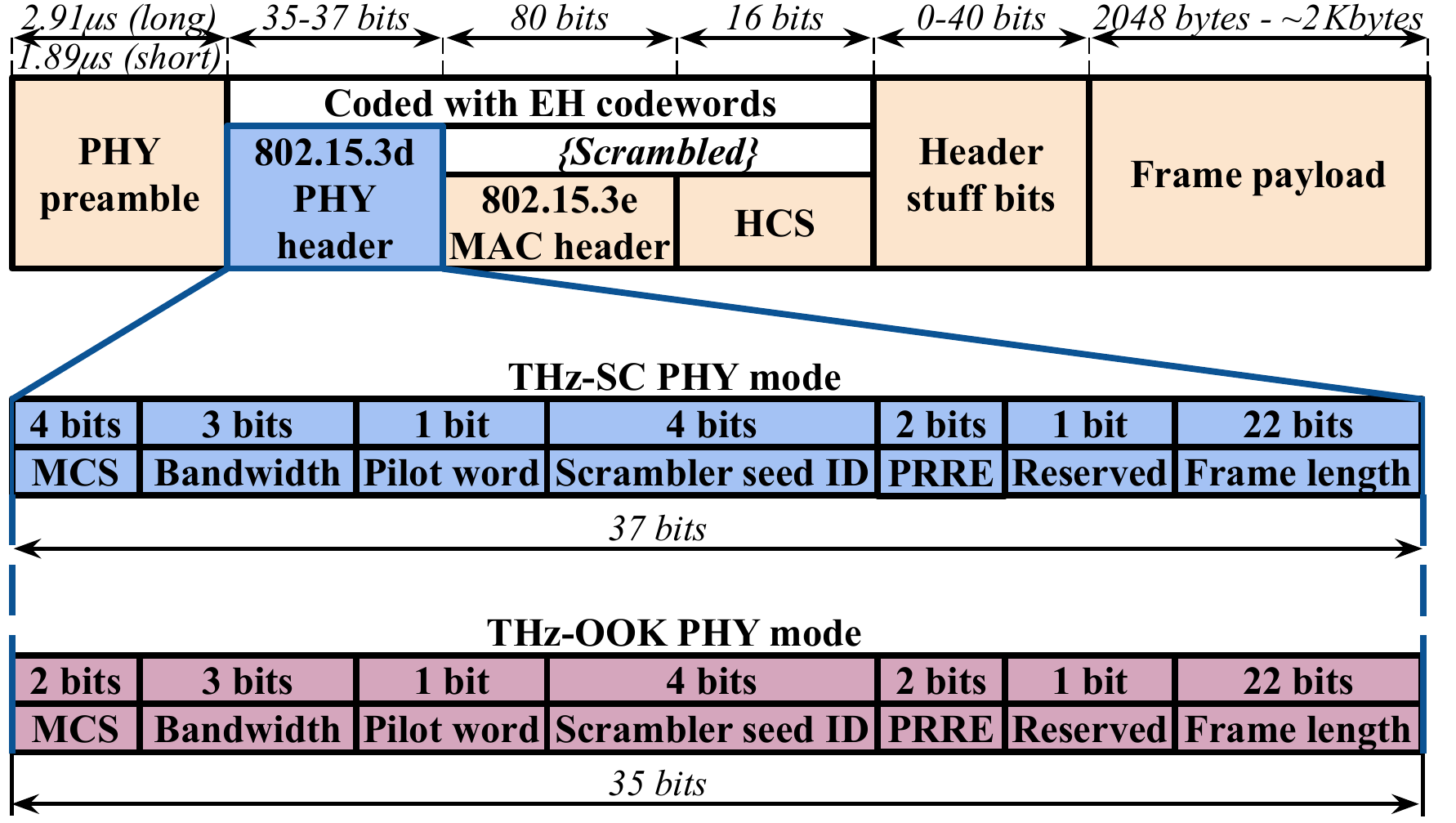}
 \caption{IEEE 802.15.3d frame format.}
 \label{fig:frame_header}
\end{figure}

\subsection{THz-SC PHY Layer Mode}
\label{sec:phy_sc}
The THz-SC mode is primarily intended for high-rate communications. According to~\cite{ieee_802_15_3d}, the frame length varies from $2048$ bytes to $2$\,$099$\,$200$ bytes, not including the PHY preamble or the base header. THz-SC PHY supports $6$ different modulations. The first four are phase shift keying modulations: binary (BPSK), quadrature (QPSK), 8-Phase (8-PSK), and 8-Phase Asymmetric (8-APSK). The mode also supports quadrature amplitude modulation with $16$ and $64$ constellation points: 16-QAM and 64-QAM. Forward error correction (FEC) is performed via one of the two low-density parity-check (LDPC) codes: either a high-rate 14/15 LDPC (1440,1344) or a low-rate 11/15 LDPC (1440,1056). BPSK and QPSK modulations are mandatory for the THz-SC mode, the others are optional.

The frame format used in the THz-SC mode is illustrated in Fig.~\ref{fig:frame_header}. The frame starts with a PHY preamble to aid frame detection, timing acquisition, and channel estimation at the receiver. There are two options for the PHY preamble: a long version is transmitted during the PSP, while the short one is transmitted during the PAP. The preamble is followed by a PHY header, detailed in Fig.~\ref{fig:frame_header}. Both the MAC and the PHY headers are protected with the header check sequence (HCS), implemented using the cyclic redundancy check (CRC), particularly, CRC-16. To increase the robustness, the combination of the PHY header, MAC header, and HCS is encoded to concatenated codewords of an extended Hamming (EH) code. Additional header stuff bits may be added to get an integer length of the data portion in the block. Finally, the optional header stuff bits are followed by the frame payload.

\subsection{THz-OOK PHY Layer Mode}
\label{sec:phy_ook}
In contrast to the THz-SC mode aiming for the highest possible data rates, the THz-OOK mode of the IEEE~802.15.3d PHY layer is intended for simpler devices. The primary use cases for this mode are kiosk downloading and intra-device communications, as described in Section~\ref{sec:applications}. The target transmission range for this mode thus does not exceed a few meters. The THz-OOK mode supports a single low-complexity modulation scheme -- on-off keying (OOK) -- where the data are represented either as the presence (logical ``1'') or absence (logical ``0'') of a carrier wave. On top of the OOK modulation, three FEC schemes are supported. The (240,224)-Reed Solomon (RS) code is mandatory for this mode, as it allows simple decoding without soft decision information. The other two FEC options are the LDPC-based schemes outlined in the previous subsection. They are \emph{optional} for the THz-OOK mode and enable the use of soft-decision information.

The frame structure used in THz-OOK mode is similar to the one used in THz-SC mode (see Fig.~\ref{fig:frame_header}) with just the minor differences. Particularly, as there is just a single modulation (OOK) plus three FEC options, only two bits are used in the PHY header to define the current modulation and coding scheme (MCS), instead of four bits in THz-SC. Other fields are identical, thus making the THz-OOK PHY header two bits shorter than the one for the THz-SC mode. The target minimal receiver sensitivity levels for these two modes are also the same and equal to $-67$\,dBm (with 11/15 LDPC code and the smallest bandwidth of $2.16$\,GHz).


\section{Initial Performance Evaluation}
\label{sec:performance}

\begin{table}[!b]
\caption{Key numerical study parameters}
\label{tab:params}
\begin{center}
\begin{tabular}{|p{0.24\columnwidth}|p{0.135\columnwidth}|p{0.125\columnwidth}|p{0.135\columnwidth}|p{0.125\columnwidth}|}
\hline
 & \textbf{Fronthaul /backhaul} & \textbf{Data \newline center} & \textbf{Kiosk download} & \textbf{Intra-device}\\
\hline
Tx power & 25 dBm & 10 dBm & 0 dBm & 0 dBm\\
\hline
Tx gain & 30 dB & 30 dB & 24 dB & 6 dB\\
\hline
Rx gain & 30 dB & 30 dB & 12 dB & 6 dB\\
\hline
\hline
Carrier frequency & \multicolumn{4}{c|}{300 GHz}\\
\hline
Rx noise figure & \multicolumn{4}{c|}{8 dB}\\
\hline
Error Vector & \multicolumn{4}{l|}{$[-22 ... -3]$ dB, depending on the selected MCS}\\
Magnitude (EVM) & \multicolumn{4}{l|}{and channel bandwidth as detailed in~\cite{thz_simulator}}\\
\hline
\end{tabular}
\end{center}
\end{table}

In this section, we report the results of the initial performance evaluation of the IEEE Std. 802.15.3d--2017.

\textcolor{black}{We first recall the maximum PHY-layer data rates reported for the IEEE Std. 802.15.3d--2017 (particularly, Table 13-4 for the THz-SC mode and Table~13-12 for the THz-OOK mode~\cite{ieee_802_15_3d}).} Point-to-point connectivity is assumed with the SNR sufficiently high to guarantee a negligible bit error rate (BER) with the selected MCS. Following~\cite{ieee_802_15_3d}, the data rates with the THz-OOK mode vary from $1.64$\,Gbit/s for the smallest band of $2.16$\,GHz up to $52.56$\,Gbit/s for the largest $69.12$\,GHz-wide channel. The more advanced THz-SC can theoretically support higher rates up to $315.39$\,Gbit/s with 64-QAM modulation and\,14/15\,LDPC\,code.

\begin{figure}[!t]
 \centering
 \includegraphics[width=1.0\columnwidth]{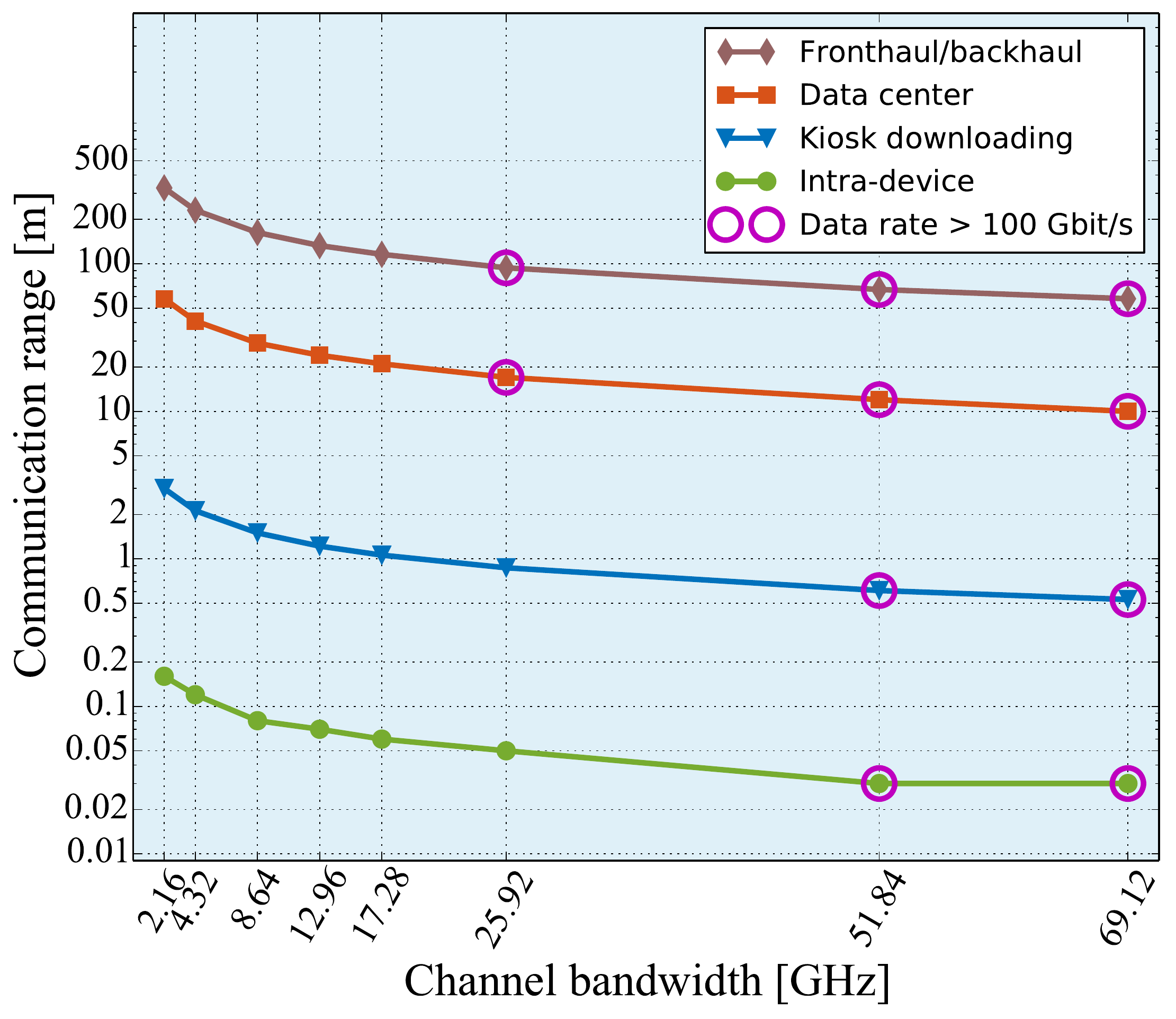}
 \caption{Range and rate values achievable in target IEEE 802.15.3d use cases.}
 \label{fig:comm_range}
\end{figure}

We then proceed with a study on the communication range, as reported in Fig.~\ref{fig:comm_range}. This figure illustrates the maximum distance between the PRC and PRDEV that still allows for reliable data exchange as a function of the channel bandwidth. For this study, a link-level simulator was developed, as detailed in~\cite{thz_simulator}. For each of the use cases, the simulator utilizes the corresponding application-specific parameters and channel models for sub-THz communication introduced in~\cite{thz_cmd}. The efficient 64-QAM modulation is used for the long-range setups (fronthaul/backhaul and data center). A more robust 8-APSK modulation is utilized for kiosk downloading and intra-device. For all the use cases, the high-rate 14/15 LDPC code is applied. The key modeling parameters are given in Table~\ref{tab:params}.

Following Fig.~\ref{fig:comm_range}, the range vary from centimeters for the intra-device and up to several hundred meters for the fronthaul/backhaul. At the same time, as detailed in~\cite{thz_simulator}, the use of highly-directional antennas is mandatory on both PRC and PRDEV sides, especially for the long-range fronthauling and backhauling. The antenna gains of $\geq30$\,dB are required for this use case to mitigate the spatial loss and the absorption. Fig.~\ref{fig:comm_range} also shows the maximum range, while still maintaining a $100+$\,Gbit/s data rate. These values reach $\approx100$\,m for the fronthaul/backhaul, $17$\,m for the data center, $0.61$\,m for the kiosk downloading, and $0.03$\,m ($3$\,cm) for the intra-device connectivity. The results are several times lower than the values achievable with a $2.16$\,GHz channel, highlighting a trade-off between the range and the link-level performance. In practical implementations, operators may adjust this range--rate balance depending on the particular application and deployment.

Summarizing, the preliminary performance assessment of the IEEE Std. 802.15.3d--2017 illustrates that the essential rate- and range-related requirements are satisfied in all the described use cases. Nevertheless, more detailed evaluations are still in progress, aiming to refine the results in specific practical setups, such as wireless backhauling in realistic environments and the data centers with complex topology~\cite{bh_detailed}.

\section{The Road Ahead in Standardizing\\THz Band Communications}
\label{sec:future}


\subsection{Regulatory Aspects of THz Band Communications in 6G}
\label{subsec:regulatory}
THz communications is one of the candidates to enable ultra-high data rates in 6G networks~\cite{rap_6g}. THz systems cannot maintain wide coverage, and therefore they will not suit as the general-purpose 6G air interface. Still, THz connections will be an important component providing extremely-high-rate data pipes. To facilitate this, an unprecedented amount of $160$\, GHz of the spectrum was recently opened for THz communications after the WRC--2019, which has included a new note RR No. FN5.564A describing the conditions to use the $275$\,GHz--$450$\,GHz band by land mobile and fixed service~\cite{wrc_final}.

One of the major issues analyzed in the regulatory process is the potential interference from the THz communications to existing ground-to-orbit links. The effect has been extensively studied over recent years. As a result, sharing with passive services is still necessary within certain parts of the specified band. In contrast, within $252$--$296$\,GHz, $306$--$313$\,GHz, $318$--$333$\,GHz, and $356$--$450$\,GHz no conditions are necessary to protect earth exploration-satellite service (EESS), and well manageable conditions exist to protect the radio astronomy service (RAS)~\cite{wrc_final}. Hence, massive bands are already available for future THz communication systems.

\textcolor{black}{The presented WRC allocations are global, they hold for the entire world and provide the baseline for further regulatory activities. No specific limits for the transmit power levels are specified to date. Additional region-specific and national regulations are expected to appear in the coming years. These regulations will follow the general agreements summarized in~\cite{wrc_final} but might specify more details, e.g., determine the allowed power levels, the allocations of the sub-bands, and the necessity and nature of the required coexistence mechanisms.
}

\subsection{Prototyping Activities for IEEE Std. 802.15.3d}
The implementation of the end-to-end communication system operating in the IEEE Std. 802.15.3d--2017 bands is currently an active work in progress. These demonstrators are primarily built within large-scale research and infrastructure projects. Particularly, the demonstrator developed within the EU-Japan project THoR applies the chipsets following the implementation in~\cite{ieee_802_15_3e}, which uses the same MCSs as defined in the THz-SC PHY mode. Another solution based on RTDs has been presented in the EU H2020 project iBROW, aiming to provide low-complexity and cost-efficient implementation that is more suitable for the THz-OOK mode.

Started several years ago, these prototyping activities have been pushed further by the final approval of the IEEE Std. 802.15.3d--2017 specification, as well as the recent frequency allocations made at the WRC-19~\cite{wrc_final}. A recent example here is a transceiver chip capable to utilize the channels 49--51 and 66 from Fig.~\ref{fig:channels}~\cite{extra_cmos_thz_tranceiver}. Further prototyping activities are upcoming, following the announced decisions and the availability of the hardware needed for sub-THz radio links.

\subsection{Towards Prospective THz Band WLANs}
The IEEE Std. 802.15.3d--2017 specifies wireless THz connections for fixed point-to-point links only, where the (i) directions of the antennas at both ends of the links are known; (ii) interference can be mitigated by proper links planning~\cite{bh_detailed}; and (iii) no fight for access exists. In prospective THz WLANs, these conditions do not hold. Therefore, the proper solutions have to be developed first to address the listed aspects.

Once scientific breakthroughs in these areas are made, the corresponding algorithms have to be included in standards. Several possibilities are currently under consideration in the IEEE 802.15 TAG THz and other teams, such as developing amendments to the existing IEEE Std. 802.15.3d--2017, incorporating selected 802.15.3d features into future releases of IEEE 802.11, or creating completely new standards. Today, the IEEE 802.15 TAG THz is also exploring novel solutions for physical and link layers to operate over $10$--$100$\,GHz-wide channels in sub-THz and THz bands. On this way forward, the ideas and solutions delivered in the IEEE Std. 802.15.3d--2017 will contribute to the appearance of THz communications, as well as form the basis for prospective 6G wireless systems.

\balance

\section*{Authors' Biographies}

\textbf{Vitaly Petrov} (vitaly.petrov@tuni.fi) is a Ph.D. candidate at the Unit of Electronics and Communications Engineering at Tampere University, Finland. He received the Specialist degree (2011) from SUAI University, St. Petersburg, Russia, and the M.Sc. degree (2014) from Tampere University of Technology. He received the Best Student Paper Award at IEEE VTC-Fall'15, Best Student Poster Award at IEEE WCNC'17, and Best Student Journal Paper Award from IEEE Finland, 2018.

\textbf{Thomas Kürner} \emph{Fellow, IEEE} (kuerner@ifn.ing.tu-bs.de) received his Dipl.-Ing. degree in Electrical Engineering in 1990, and his Dr.-Ing. degree in 1993, both from Univerity of Karlsruhe (Germany). Since 2003, he is a Full University Professor at the Technische Universität Braunschweig, working on simulation and modeling of mobile radio systems, including cellular networks, THz communications, and V2X communications. He was the chair of the IEEE 802.15.3d Task Group. Currently, he is chairing the IEEE 802.15 TAG THz, is the EU coordinator of the EU-Japan Project ThoR and the spokesman of the DFG Research Unit Meteracom on Metrology for THz Communications. He received the Neal Shepard Award, 2019.

\textbf{Iwao Hosako} \emph{Member, IEEE} (hosako@nict.go.jp) received the B.S., M.S., and Ph.D. degrees from the University of Tokyo, Tokyo, Japan, in 1988, 1990, and 1993, respectively. He is currently Director General of the Wireless Networks Research Center at NICT, Japan. His research is focused on terahertz band devices and systems. Dr. Hosako is serving on several research committees on terahertz technologies. He was the vice-chair of the IEEE 802.15.3d Task Group, and currently he is the vice-chair of the IEEE 802.15 TAG THz.

\end{document}